\documentstyle[twoside,fleqn,espcrc2,epsf]{article}

\newcommand{\la}{\raise.16ex\hbox{$\langle$}}
\newcommand{\ra}{\raise.16ex\hbox{$\rangle$}}
\newcommand{\be}{\begin{equation}}
\newcommand{\ee}{\end{equation}}
\newcommand{\bea}{\begin{eqnarray}}
\newcommand{\eea}{\end{eqnarray}}

\title{Static three- and four-quark potentials}

\author{
C.~Alexandrou~$^a$~\thanks{Talk presented by C.~Alexandrou.},
Ph.~de~Forcrand~$^{b,c}$,
and  
A.~Tsapalis~$^d$~\thanks{Acknowledges funding from the 
University of Cyprus and the European network ESOP (HPRN-CT-2000-00130).}\\
\vspace{3mm}
$^a$~Department of Physics, University of Cyprus, 
CY-1678 Nicosia, Cyprus\\
$^b$~Institut f\"ur Theoretische Physik, ETH-H\"onggerberg, CH-8093 Z\"urich,
Switzerland\\
$^c$~Theory Division, CERN, CH-1211 Geneva 23, Switzerland\\
$^d$~Department of Physics, University of Athens, Athens, Greece
}

\begin{document}

\begin{abstract}
We present results for
 the static three- and four- quark potentials in $SU(3)$ and $SU(4)$
respectively. Using a variational approach, combined with multi-hit 
for the time-like links, we determine the ground state of the baryonic string
with sufficient accuracy to test the $Y-$ and $\Delta-$ ans\"atze for the baryonic 
Wilson area law. 
Our results favor the $\Delta$ ansatz, 
where the potential is the sum of two-body terms.
\end{abstract}

\maketitle

\begin{flushleft}
\vspace*{-10cm}
CERN-TH/2001-280
\vspace*{9.2cm}
\end{flushleft}

\vspace*{-1.5cm}

\section{Introduction}
There are many lattice studies of the $q\bar{q}$ potential
emphasizing the  important role it plays in our understanding
of  the structure of mesons.
Despite the equally important role that the 
three-quark potential plays in the
understanding of baryon structure 
 it has, until recently, received little attention in lattice QCD studies. 
Now, two lattice studies of the three quark potential have appeared during the
last year \cite{Bali,Takahashi},
which reach different conclusions for the area law behaviour of the
baryonic Wilson loop.
It must be stressed that the main difficulty to resolve the dominant area law 
for the baryonic potential is the fact that the maximal
difference between the two ans\"atze 
is a mere  15\% for $SU(3)$.
In this work we reexamine the baryonic potential using state of the
art lattice techniques~\cite{AFT}. 
Since  the
same issues arise for any gauge group $SU(N)$
we corroborate our conclusions by also studying  $SU(4)$, choosing
lattice geometries which maximize the difference between the
two ans\"atze to the 20\% level.

\section{Baryon Wilson loop}
In $SU(N)$ we create a gauge invariant 
$N$-quark state at time $t=0$ which is annihilated at a later time $T$.
In $SU(3)$
the baryon Wilson loop $W_{3q}$, shown in Fig.~1, is given by
\small
\be
\frac{1}{3!}\epsilon^{abc}\epsilon^{a'b'c'} U({\bf x,y},1)^{aa'}
        U({\bf x,y},2)^{bb'}U({\bf x,y},3)^{cc'}
\ee
 
\normalsize
\noindent
where 
\be
U(x,y,j)=P\exp\left[ig\int_{\Gamma(j)} dx^{\mu}A_\mu(x)\right] \quad,
\ee
$P$ is the path ordering and $\Gamma(j)$ denotes the path from $x$ 
to $y$ for quark line $j$.

The $N$-quark potential is then extracted
from the long time behaviour of the Wilson loop:
\be
V_{Nq}=-\lim_{T \rightarrow \infty} \frac{1}{T} \ln<W_{Nq}> \quad.
\ee


\begin{figure}[h]
\vspace*{-0.8cm}
\epsfxsize=5.5truecm
\epsfysize=5.5truecm
\mbox{\epsfbox{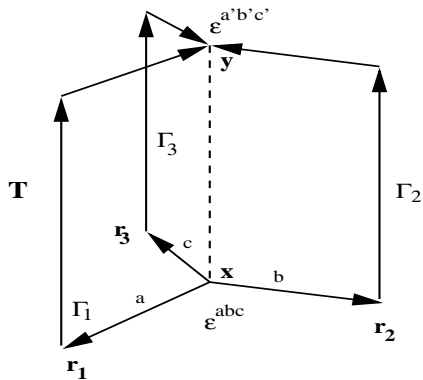}}
\vspace*{-0.8cm}
\caption{The baryonic Wilson loop in $SU(3)$. 
The quarks are located at positions ${\bf r}_1,{\bf r}_2$ and ${\bf r}_3$.}
\vspace*{-1.cm}
\end{figure}

\section{Geometries in $SU(3)$ and $SU(4)$}
\begin{figure}[h]
\epsfxsize=6.0truecm
\epsfysize=5.truecm
\mbox{\epsfbox{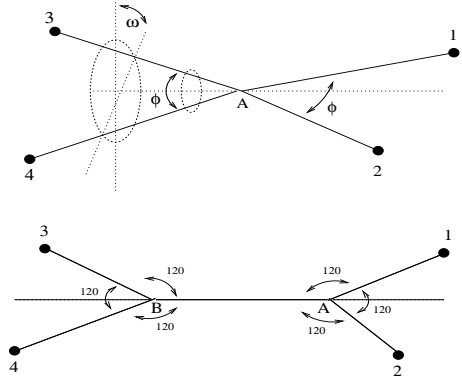}}
\vspace*{-0.8cm}
\caption{Flux tube configurations for four quarks
at positions ${\bf r}_1,{\bf  r}_2, {\bf r}_3$ and ${\bf r}_4$. 
The upper graph ($X$) shows the local minimum of the energy
 with one Steiner point A, and
the lower ($Y$) is the minimum with two Steiner points A and B.}
\label{fig:BWL su4}
\vspace*{-0.5cm}
\end{figure}

Two ans\"atze exist in the literature regarding the area law behaviour
of the baryon Wilson loop:
\begin{itemize}
\item 
$Y-$ansatz:
In the strong coupling limit,
 minimization of 
the static energy gives the flux tubes of shortest total length  $L_Y$
 joining the quarks.
For $SU(3)$,
if the three quarks are at positions ${\bf r_1},{\bf r_2}$ and
${\bf r_3}$,
the flux tubes in that configuration will meet at an interior point~\cite{CKP},
known as the Steiner point, where the angles between the flux tubes
are $120^{0}$
independently of the vectors ${\bf r}_k$, provided that none of the angles
of the triangle formed by the three quarks exceeds $120^{0}$.
[Otherwise, the configuration of minimal total length is made
of two flux tubes meeting at the third quark location.]
Time evolution of the general case produces 
a three-bladed area similar to Fig.~1, known as the $Y-$ area law.

For $SU(4)$,
minimization of the static energy leads to two possibilities
as shown in Fig.~\ref{fig:BWL su4}: namely,
 one configuration with two Steiner points, A and B ($Y-$ansatz) and one  with a 
single Steiner point ($X-$ansatz).
If, for simplicity,  we assume that the double string between the 
two Steiner points has the same tension as the other four, single strings, then 
the $Y-$ansatz always has lower energy.
In fact, the tension of the double string is 1.357(29) times greater 
\cite{Teper2}. Since this further increases
the potential of the $Y$-ansatz, it will turn out to have no bearing on our
conclusions. 
The two Steiner points
are obtained by an iterative numerical procedure. 

\item 
$\Delta$-ansatz:
The second possibility~\cite{Cornwall} for the relevant area 
dependence of the baryonic Wilson loop is that it is given by the sum
of the minimal areas $A_{ij}$ spanning quark lines $i$ and $j$ which
because of its shape in $SU(3)$ is known as
the $\Delta$- area law, with $L_{\Delta}$ the length  of all
interquark distances.
The potential is then a sum of two-body potentials.

\end{itemize}

For $SU(3)$, the maximal difference of 15\% between the two proposed area laws
is obtained when the three
 quarks form an equilateral triangle.

For $SU(4)$, it
turns out that the relative difference between the $Y-$ energy 
and the two-body law
is also 
maximal for the configuration of maximal symmetry among the four quarks.
This entails putting the quarks on the vertices of a regular
tetrahedron, which gives a relative difference of $\approx 22 \,\, \% $ 
between the two ans\"atze.
We make no attempt here to distinguish between the
$Y$- and $X$- ans\"atze, since for the highly symmetric geometries that
we choose the difference is on the few percent level.

The lattice results are thus compared to the 
two expected forms of the baryonic potential which in  $SU(N)$  are

\vspace*{-0.3cm}

\be
\hspace*{-0.3cm}
V_{Nq}({\bf r_1,\cdots,r_N}) = 
\frac{N}{2} \!V_0 
       -\frac{1}{N-1}\sum_{j<k}\frac{g^2 C_F}{4\pi r_{jk}} 
+ \sigma \left\{ \!\!\begin{array}{c} \frac{L_\Delta}{N-1} \\ L_Y \end{array}
                     \! \!\right\} 
\ee

\vspace*{-0.2cm}

\noindent
with $C_F=(N^2-1)/2N$ 
and $\sigma$ the string tension of the $q\bar{q}$ potential. 
The factor of 1/(N-1) in the $\Delta-$ ansatz makes
$L_\Delta/(N-1) < L_Y$.
Note that, in contrast to Ref.~\cite{Takahashi}, we do not vary $\sigma$.   
In addition, the three- or four-quark potential is compared directly with the sum of
two-body potentials measured on the same gauge configurations, with no
adjustable parameters.

\section{Lattice techniques}
We use the
multi-hit procedure for the time-like links.
For $SU(3)$, the mean-link integral is obtained
analytically,
whereas for $SU(4)$ a Monte Carlo integration is performed~\cite{AFT}.
In addition, we use a variational approach, 
and consider $M$ different levels of APE smearing for the spatial links, 
with optimized parameters as in ref.~\cite{alpha}.
Therefore, we obtain an $M\times M$ correlation matrix $C(t)$ of Wilson loops.
To obtain the groundstate potential, we solve the generalized eigenvalue problem
$C(t)v_k(t)=\lambda_k(t)C(t_0)v_k(t)$, taking $t_0/a=1$. 
In the first variant the potential levels are  extracted via
$
aV_k={\rm Lim}_{t\rightarrow \infty} -\ln
\left(\frac{\lambda_k(t+a)}{\lambda_k(t)}\right)
$
by fitting to the plateau.
In the second variant we consider the projected Wilson loops 
$
W_P(t)=v_0^T(t_0)C(t)v_0(t_0)
$
and fit $aV_0$ to the plateau value of  
$-{\rm ln}\biggl(W_P(t+1)/W_P(t)\biggr)$.
Both procedures give consistent
results.
The energy for the
first excited state was used to check, in the extraction of 
the ground state, that the
contamination is less than $e^{-2}$.

\section{Results}

For the baryonic loop in $SU(3)$ we
 used 220 configurations at $\beta=5.8$ and 200 at $\beta=6.0$ for a lattice 
of size $16^3\times 32$ from the NERSC archive.
For $SU(4)$ we generated 100 configurations at $\beta=10.9$,
which gives a similar string tension $\sigma a^2$ as $SU(3), \beta=6.0$.

\begin{figure}[h]
\vspace*{-2.5cm}
\epsfxsize=5.5truecm
\epsfysize=7.truecm
\mbox{\epsfbox{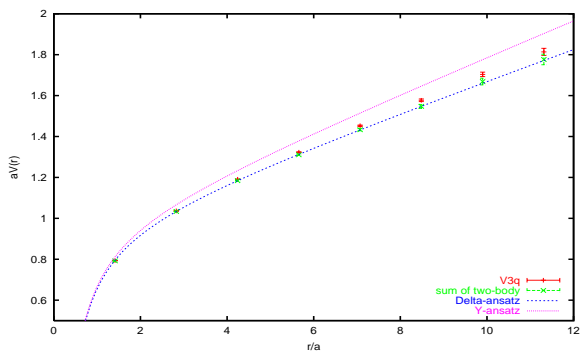}}
\vspace*{-1cm}
\caption{The static baryonic potential at $\beta=6.0$ (pluses). The crosses
show the sum of the static $q\bar{q}$ potentials. The curves for the 
$\Delta$ and $Y$ ans\"atze are also displayed. The quarks are located
at $(l,0,0)$, $(0,l,0)$, $(0,0,l)$, and $r = l \sqrt{2}  $.}
\label{fig:beta60}
\vspace*{-0.5cm}
\end{figure}

\begin{figure}[h]
\vspace*{-2.5cm}
\epsfxsize=5.5truecm
\epsfysize=7.truecm
\mbox{\epsfbox{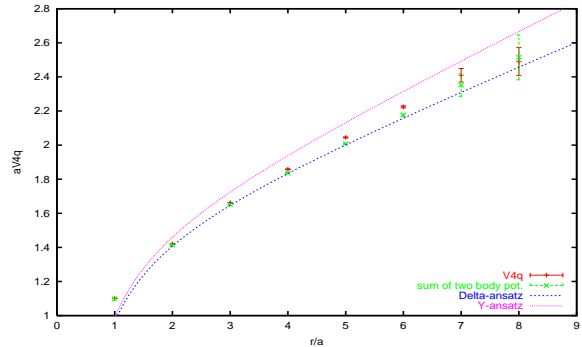}}
\vspace*{-1cm}
\caption{Same as in Fig.\ref{fig:beta60}, for the static $SU(4)$ baryonic 
potential. The quarks
are located at $(r,0,0)$, $(0,r,0)$, $(0,0,r)$ and $(0,0,0)$.}
\label{fig:su4 geometry 3}
\vspace*{-0.5cm}
\end{figure}

\section{Conclusions}
Our results for the static three- and four-quark potentials
in $SU(3)$ and $SU(4)$ are consistent with the sum of two-body potentials
up to a distance of about $0.8$~fm,
and  inconsistent with the $Y-$ ansatz.
For larger distances, where our statistical and systematic errors both
become appreciable, there appears to be a small enhancement due to
an admixture of a many-body component. 
Nevertheless, for the distances
up to 1.2~fm that we were able to  probe in this work, 
the $\Delta$ area law gives the closest description of our data. 
More refined noise-reduction techniques, such as the L\"uscher-Weisz~\cite{LW}
algorithm, will be needed
in order to clarify whether a genuine many-body component is present
at large distances.

\end{document}